\documentclass[desactivate]{aa}  

\usepackage{graphicx}
\usepackage{txfonts}
\usepackage{hyperref}
\hypersetup{ colorlinks, linkcolor=blue, citecolor=blue }

\defcitealias{Noll2024}{N24}

 \newcommand{\carbon}{\element[][12]{C}}
\newcommand{\oxygen}{\element[][16]{O}} \newcommand{\ov}{\alpha_{\mathrm{ov}}}
 
\newcommand{\dd}{\mathrm{d}} 
\newcommand{\dex}{\mathrm{dex}} 
 \newcommand{\smass}{M_{\odot}}

 \newcommand{\gradad}{\nabla_{\rm
ad}} \newcommand{\gradrad}{\nabla_{\mathrm{rad}}}
\newcommand{\dov}{d_{\mathrm{ov}}} \newcommand{\brunt}{Brunt-Väisälä }
\newcommand{\reac}{\carbon(\alpha,\gamma)\oxygen}

\title{Ensemble seismic study of the properties of the core of Red Clump stars}

\author{Anthony Noll\inst{1,2,3}, Sarbani Basu\inst{4}, and Saskia
Hekker\inst{3,5}} \institute{Institut de Ciencies de l'Espai (ICE, CSIC), Carrer
de Can Magrans S/N, 08193 Bellaterra, Spain\label{1} \and Institut d’Estudis
Espacials de Catalunya (IEEC), Carrer Gran Capita 4, 08034 Barcelona,
Spain\label{2} \and Heidelberger Institut für Theoretische Studien,
Schloss-Wolfsbrunnenweg 35, 69118 Heidelberg, Germany\label{3} \and Department
of Astronomy, Yale University, PO Box 208101, New Haven, CT 06520-8101,
USA\label{4} \and Center for Astronomy (ZAH/LSW), Heidelberg University,
Königstuhl 12, 69117 Heidelberg, Germany\label{5}}


\begin{document}
\abstract{Red clump (RC) stars still pose open questions regarding several physical processes, such as the mixing around the core, or the nuclear reactions, which are ill-constrained by theory and experiments. The oscillations of red clump stars, which are of mixed gravito-acoustic nature, allow us to directly investigate the interior of these stars and thereby better understand their physics. In particular, the measurement of their period spacing is a good probe of the structure around the core.}{We aim to explain the distribution of period spacings in red clump stars observed by \emph{Kepler} by testing different prescriptions of core-boundary mixing and nuclear reaction rate.}{Using the MESA stellar evolution code, we computed several grids of core-helium burning tracks, with varying masses and metallicities. Each of these grids have been computed assuming a certain core boundary mixing scheme, or $\reac$ reaction rate. We then sampled these grids, in a Monte-Carlo fashion, using observational spectroscopic metallicities and seismic masses priors, in order to retrieve a period spacing distribution that we compared to the observations.}{We found that the best fitting distribution was obtained when using a ``maximal overshoot'' core-boundary scheme, which has similar seismic properties as a model whose modes are trapped outside a semi-convective region, and which does not exhibit core breathing pulses at the end of the core-helium burning phase. If no mode trapping is assumed, then no core boundary mixing scheme is compatible with the observations. Moreover, we found that extending the core with overshoot worsens the fit. Additionally, reducing the $\reac$ reaction rate (by around $15\%$) improves the fit to the observed distribution. Finally, we noted that an overpopulation of early RC stars with period spacing values around 250\,s is predicted by the models but not found in the observations.}{Assuming a semi-convective region and mode trapping, along with a slightly lower than nominal $\reac$ rate, allows us to reproduce most of the features of the observed period spacing distribution, except for those of early RC stars.}
\maketitle

\section{Introduction}
The red clump (RC) is composed of low-mass (approximately less than $1.8\,\smass$) stars that have gone through the helium-flash and are currently in the core helium burning (CHeB) stage. An interesting property of these RC stars is that their luminosity and, more generally, the properties of their core, depend little on the mass of the star nor on pre-core helium burning evolution.  Therefore, they can be used as standard candles and/or tracers of the evolution of the composition of the Galaxy (see the review by \citealt{Girardi2016}).  

Yet, the structures of RC stars are not fully understood, due to the lack of theoretical prescriptions for several physical processes. In particular, the properties of the core boundary mixing (CBM) are ill-defined by theory \citep{Castellani1971b,Castellani1971,Bressan1986}. Furthermore, the nuclear reaction rate of the $\reac$ reaction, despite progress in the last decades, is still subject to significant uncertainties \citep{Kunz2002,deBoer2017,Shen2023}. 

Asteroseismology provides a way to put observational constraints on these physical processes. RC stars are solar-like oscillators, i.e., their modes are excited by the turbulent motion of fluid in the convective envelope. Moreover, their non-radial modes are mixed: they propagate as gravity waves in a region contained around the convective core and as pressure waves in the outer part of the star. Because of this, these modes are particularly sensitive to the properties of the region around the convective core. A key aspect of the mixed modes of evolved post-main sequence stars, like the RC stars, is that they closely follow an asymptotic relation \citep{shibahashi79}. This allows us to define a period spacing, $\Delta \Pi$, that is a direct probe of the region around the convective core. Thanks to the data from the \emph{Kepler} satellite \citep{borucki10}, the period spacing of thousands of RC stars has been measured \citep{Mosser2012,Mosser2014,Vrard2016}, which opened a new window on the internal properties of these stars. Notably, \cite{Montalban2013} showed that extending the convective core beyond the boundary defined by the Schwarzschild criterion is necessary to reproduce the observations. The question of the nature of the CBM has been investigated by several works, with varying results: {\it ad hoc} ``maximal'' extension of the core \citep{Constantino2015}, mild extension of the core with a radiative \citep{Bossini2015} or adiabatic \citep{Bossini2017} temperature stratification. Moreover, \cite{Noll2024} (\citetalias{Noll2024} hereafter), found that a straight-forward core extension, such as overshooting or penetrative convection, could not explain the seismic observations. This result, which differs from \cite{Bossini2015,Bossini2017}, is due to the fact that \citetalias{Noll2024} use a different algorithm to determine the convective boundaries.  Specifically, they use the convective premixing scheme, which ensures convective neutrality at the boundaries, following the recommendations of \cite{gabriel14}.  Furthermore, similarly to \cite{Constantino2015,Constantino2017}, \citetalias{Noll2024} raised the possibility that the observed modes could be trapped outside the CBM region, which would impact the observed period spacing. Additionally, these authors showed that the rate of the $\reac$ reaction, when increased, lengthens the duration of the core-helium burning phase such that the maximum value of period spacing reached by the models increases. All these investigations show the strong potential of asteroseismology, and in particular the study of the period spacing, to better understand the physics of RC stars.

In this work, we performed an ensemble study of the period spacings of a sample of RC stars observed by \emph{Kepler}. To do so, we simulated the observed distribution of period spacings, taking into account the variations of metallicity and mass within the sample and using different prescriptions of the physics of the models. The aim is to test the validity of these prescriptions, by comparing the distribution of the models' period spacings with the observed one. In Sect.~\ref{section_method}, we introduce the method as well as the properties of the models used in this work. We then show, in Sect.~\ref{section_results}, how the distributions that we obtained assuming different physics compare with the observed one. Next, in Sect.~\ref{the_250s_peak}, we discuss the so-called 250\,s peak, that is the largest difference between the modeled distribution and the observations. Finally, we conclude in Sect.~\ref{section_conclusion}.

\section{Method}
\label{section_method}

In this work, we performed Monte-Carlo simulations of the period spacing distribution observed by the \emph{Kepler} satellite, and whose properties are described in Section~\ref{observations}. To do so, for each physical assumption (on core boundary mixing, or nuclear reaction rates), we computed a grid of models with varying masses and metallicities. The properties of these grids are presented in Sect.~\ref{grid_properties}. We then randomly sampled these grids of tracks using age, mass and metallicity priors that are similar to the ones of the \emph{Kepler} sample, as detailed in Sect.~\ref{grid_sampling}. The resulting period spacing distribution is finally compared to the observations.

\subsection{Properties of the observed sample}
\label{observations}

\begin{figure}
    \centering
    \includegraphics[width=9cm]{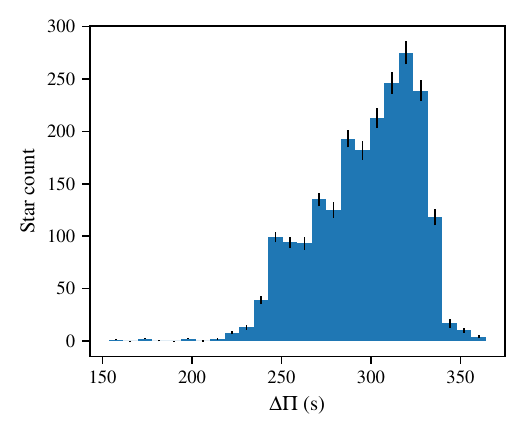}
    \caption{Period spacing distribution of the observed sample (data from
    \citealt{Vrard2016}). The uncertainties are computed following the procedure
    described in Appendix~\ref{uncertainty_bin}.}
    \label{hist_obs}
\end{figure}

The observational sample that we use in this work are taken from the cross-match of the data from the APOKASC-3 catalogue \citep{Pinsonneault2025} and \cite{Vrard2016}. The period spacings values are taken from \cite{Vrard2016}, the masses from the seismic values of \cite{Pinsonneault2025} and the metallicities and alpha-enrichment values from the spectroscopic data of the same work, which are taken from APOGEE DR16 and DR17 \citep{Ahumada2020,Abdurro'uf2022}. The metallicity values have been altered to take into account the effect of the $\alpha$-enrichment, following the prescription from \cite{Salaris1993}, with coefficients that are recomputed in order to match with the \cite{gs98} mixture used in the models: $[\mathrm{Fe}/\mathrm{H}]_{\mathrm{mod}} = [\mathrm{Fe}/\mathrm{H}]_{\mathrm{obs}} + \log \left(0.683 \times 10^{[\alpha/\mathrm{Fe}]} + 0.320 \right)$.

From this sample, we only consider stars whose values of mass and metallicity are covered by our grids of models (see Sect.~\ref{grid_properties}): masses between $0.8$ and $1.8\,\smass$, and metallicities between $-0.8\,\dex$ and $0.4\,\dex$.  Furthermore, we apply an uncertainty cut and only include stars for which the quoted uncertainty on the period spacing is smaller than $6\,\mathrm{s}$. The final sample consists of 2110 stars.

Figure~\ref{hist_obs} shows the observed period spacing distribution. The uncertainties of the bin counts are computed taking into account the observational uncertainties, following the approach explained in Appendix \ref{uncertainty_bin}. 

\subsection{Properties of the models}

\subsubsection{Microphysics}

\label{grid_properties}

The models used in this work have been computed with the MESA stellar evolution code, revision 22-11.1 \citep{Paxton2011,Paxton2013,paxton15,paxton18,Paxton2019,Jermyn2023}, with physics similar to \citetalias{Noll2024}. The opacities are computed using the OPAL code \citep{opal_opacities}. The equation of state is a mixture of FreeEOS \citep{Irwin2012} and Skye \citep{Jermyn2021}.  The convection model comes from \cite{Kuhfuss1986}. The mixing-length parameter has been fixed to 1.8, and is not varied as it does not impact the properties of the core and therefore the period spacing. The nuclear reaction rates are from the REACLIB database \citep{Cyburt2010}, and in particular from \cite{Xu2013} for the $\reac$ reaction. We took care of taking into account all the reactions of the pp-chain, as recommended by \cite{Noll2023}. Finally, we use the solar mixture from \cite{gs98}.

\subsubsection{Core boundary mixing}

\begin{figure*}
    \centering
    \includegraphics[width=\linewidth]{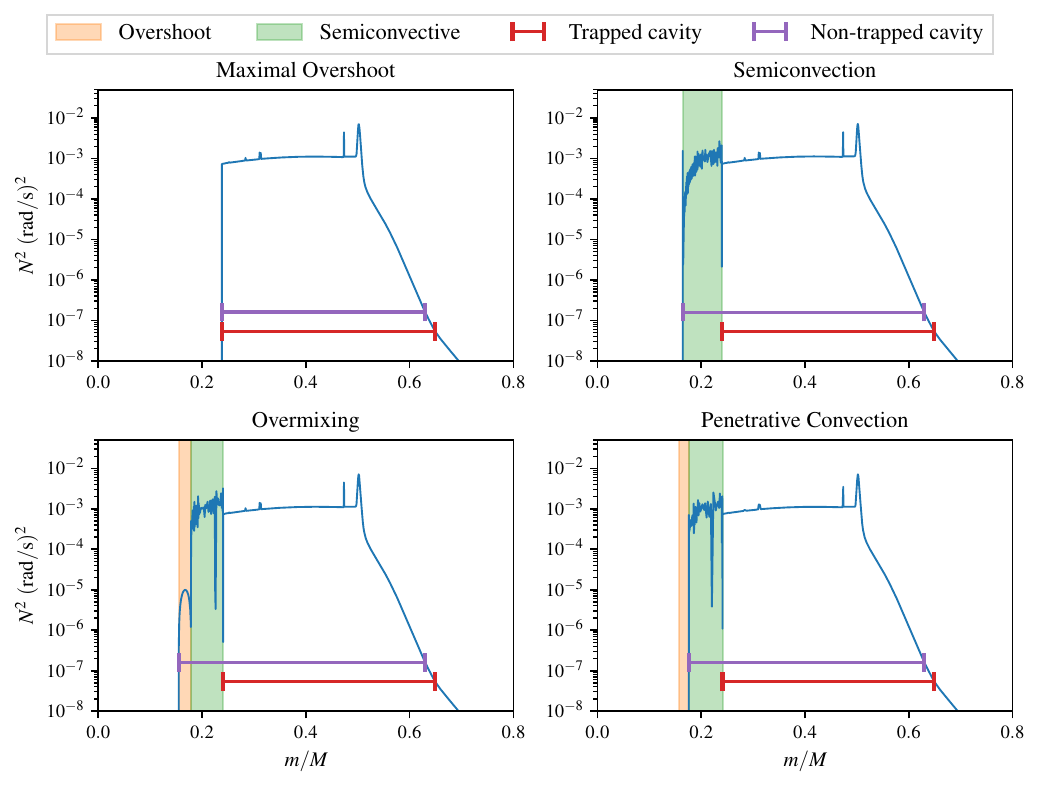}
    \caption{Brunt-Väisälä profiles of 4 models with different CBM schemes, all stopped at $Y_{\mathrm{c}} = 0.3$. We indicate in orange the overshoot region, that is fully chemically mixed with the convective core. The region in green indicates the semiconvective region, where $\gradrad = \gradad$. Finally, we indicate the region over which $N/r$ is integrated in the trapped mode scenario (red) and in the non-trapped mode scenario (purple).}
    \label{structure_models}
\end{figure*}

In this section, we briefly describe the four core boundary mixing assumptions that we test in this work. For more details, we refer the reader to \citetalias{Noll2024}. 

\paragraph{Semiconvection:} the region outside the fully-mixed part of the core is semiconvective, i.e. the composition is such that $\gradrad = \gradad$, $\gradrad$ and $\gradad$ being the radiative and adiabatic gradients, respectively. The presence of such a region in CHeB stars has been first predicted by \cite{Schwarzschild1969} and studied more thoroughly in \cite{Castellani1971b,Castellani1971}. As its definition differs from a parametrized semi-convection model, like the one of \cite{Langer1983}, it is often referred as induced semiconvection. In this work, we obtain such semiconvective regions using the convective premixing scheme \citep{Paxton2019}.  We note that for many tracks with semiconvection included we find core breathing pulses (CBP) at the end of the CHeB phase. More details on this process are given in Sect.~\ref{handling_cbp}.

\paragraph{Overmixing:} the convective core is extended over a distance $\dov = \ov H_p$, with $\ov$ a free parameter and $H_p$ the pressure scale height. The temperature gradient is taken as $\nabla = \gradrad$ in the overshoot region, with $\nabla$ being the temperature gradient. Since we use the convective premixing scheme to determine the convective boundaries, a semiconvective region occurs around the overshoot region for the more evolved models, as described in \citetalias{Noll2024}. We note that, in order to ensure that $\gradrad = \gradad$ in the semiconvective region beyond the overshoot region, we added a supplementary call to the convective premixing routine in MESA, done after the burning.\footnote{We noted that this supplementary call could lead to additional CBPs at the very end of the CHeB phase. Thus, we deactivated it in these cases.} Overmixing models exhibit CBPs at the end of the CHeB phase. To suppress these, we neglected the gravitational energy term in the energy equation, following the recommendations of \cite{Dorman1993} (see Sect.~\ref{handling_cbp}). 

\paragraph{Penetrative convection:} this scheme is similar to the overmixing scheme, but the temperature gradient is set to $\nabla = \gradad$ in the overshooting region. For penetrative convection models as well, we neglected the gravitational energy term in the energy equation. 

\paragraph{Maximal overshoot (MO):} once a local minimum appears in the radiative gradient profile, we define the core size such that this local minimum is equal to the adiabatic gradient. This scheme, which was introduced in \cite{Constantino2015}, is by definition non-physical: the value of the radiative gradient at the outer boundary of the mixed region is significantly larger than the adiabatic gradient. However, as shown in Appendix~\ref{appendix_eq_mo_sc}, it gives similar period spacing as a model with an induced semi-convective region and observed modes that are trapped outside of the semi-convective region. Yet, contrary to the semi-convective models, models computed with maximal overshoot do not show CBPs.  

\subsubsection{Handling the CBPs}
\label{handling_cbp}

\begin{figure}
    \centering
    \includegraphics[width=9cm]{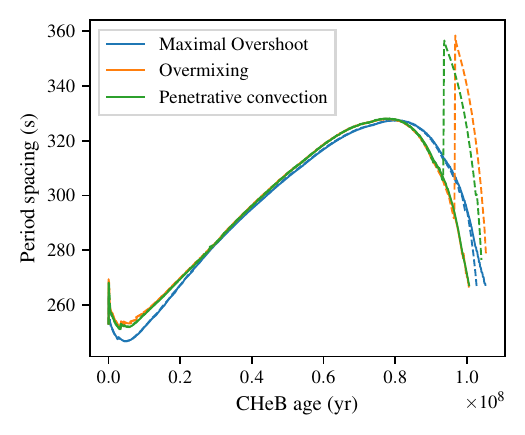}
    \caption{Evolution of the period spacing during the CHeB phase for models computed with (dashed line) and without taking into account $\epsilon_g$ in the energy equation (full line).}
    \label{evol_dpi_noepsg}
\end{figure}

CBPs are sudden increases of the convective core size that occur at the end of the CHeB phase. They are caused by the fact that, when the central helium abundance $Y_{\mathrm{c}}$ is low, a small intake of helium results in a large relative variation of the helium abundance in the core, which leads to a significant increase of the energy production and hence of the core size \citep{Sweigart1972}. CBPs seem to be a numerical artifact rather than an actual instability happening inside the stars. Indeed, their occurrence depends on the precision of the grid or timestep \citep{Dorman1993}, and they are suppressed when using a non-local implementation of the mixing \citep{Bressan1986} or when taking into account a maximum helium ingestion rate \citep{Spruit2015,Constantino2017}.  Moreover, observations of the ratio between asymptotic giant branch and horizontal branch stars in clusters \citep{Caputo1989,Constantino2016} and of the period spacing in asteroseismology \citep{Constantino2015,Noll2024} are not aligned with predictions coming from models that include CBPs. 

Therefore, we decided to suppress CBPs in models that include overshoot. To do so, we followed the approach of \cite{Dorman1993}, in which the authors force the models to be at thermal equilibrium by neglecting $\epsilon_g \equiv -T \dd s / \dd t$, with $T$ the temperature, $s$ the specific entropy and $t$ the time. The reason for that is that the extension of the core, caused by the increase in the energy production, leads to a negative $\epsilon_g$ at the core boundary, which in turn extends the core even more, causing a runaway extension. By neglecting this term, the CBPs are rapidly damped. 

To investigate how neglecting $\epsilon_g$ affects the evolution of period spacings, we compute the evolution of $\Delta \Pi$ for different CBM, with and without neglecting $\epsilon_g$. The evolution of the period spacing for these models is shown in Fig.~\ref{evol_dpi_noepsg}. One can see that, models computed without taking into account $\epsilon_g$ do not exhibit core breathing pulses at the end of the CHeB phase, while the effect on the rest of the CHeB phase is negligible. Moreover, we added the evolution of $\Delta \Pi$ for a maximal overshoot model. Even though we consistently include $\epsilon_g$ in the computation of these models in the rest of the work, it serves here as a reference case to investigate the effect of neglecting $\epsilon_g$. Thus, we find that it has little impact during most of the CHeB phase except for the very end, during the core contraction. These differences occur during a small fraction of the CHeB phase, such that the final effect on the simulated $\Delta \Pi$ distributions is small. Finally, we got rid of the residual CBPs if they happened at the very end of the helium burning phase ($Y_c < 0.015$).

\subsection{Properties of the grids and sampling priors}
\label{grid_sampling}

\begin{figure}
    \centering
    \includegraphics[width=9cm]{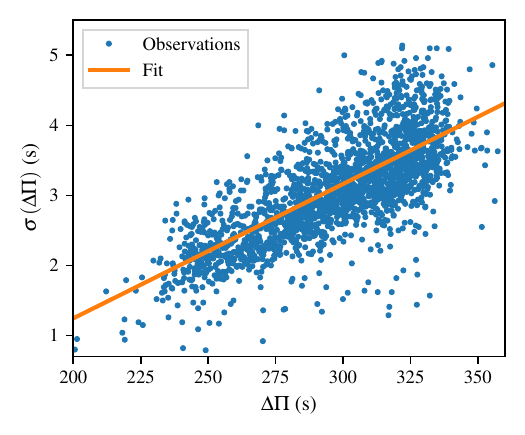}
    \caption{Period spacing uncertainties of the sample of \cite{Vrard2016}, plotted against
    the period spacing values of the same work. The orange line shows the linear
    fit that we use to simulate the uncertainties in our simulations.}
    \label{uncert_fit}
\end{figure}

The models in each of the grids have been computed to have masses between $0.8$ and $1.8\,\smass$, in steps of $0.2\,\smass$, and with metallicities $[Z/X]_0$ between $-0.8$ and $0.4\,\dex$, in steps of $0.2\,\dex$. The initial helium abundance $Y_0$ is computed using a commonly used enrichment law $Y_0 = 0.24 + 2\,Z_0$. 

We sample the period spacings of these tracks in a Monte-Carlo fashion, with mass, metallicity and age priors being as close as possible to the ones of the observed sample. Regarding age, we define $\tau$ a ``normalized'' CHeB age, that is an affine transformation of the age of the star such that $\tau = 0$ at the start of the CHeB phase, and $\tau = 1$ at the end. Our prior in $\tau$ is then uniform, between 0.01 and 0.99. For the mass and the metallicities, we use the seismic and spectroscopic observation distributions from \cite{Pinsonneault2025}, respectively. The metallicity values have been altered to take into account the $\alpha$-enrichment, as explained in Sect.~\ref{observations}. We then perform an inverse transform sampling to obtain a set of masses and metallicities values that have the same distribution as the observational priors. 

To obtain the period spacings corresponding to the values of $\tau$, masses and metallicities randomly drawn as described above, we perform 3-D linear interpolations of the period spacing within the grid. To simulate the observational period spacing uncertainties, we perturb the values obtained through the interpolation by adding a realization of $\mathcal{N} \left[0,\sigma^2(\Delta \Pi)\right]$, $\mathcal{N}$ being the normal distribution. We model the uncertainty $\sigma(\Delta \Pi)$ as a linear function of $\Delta \Pi$, fitted to the actual observational uncertainties, as showed in Fig.~\ref{uncert_fit}. This allows us to take into account the correlation between the uncertainties and the period spacing. Moreover, in the case of models computed in the ``non-trapped'' scenario (see Sect.~\ref{subsub_period_spacings}), the evolution of the period spacing is slightly noisy (see, e.g., Fig. 4 of \citealt{Noll2024}). Thus, for these models, we added in quadrature a numerical noise of $1\,\mathrm{s}$. 

For each test presented in Sect.~\ref{section_results}, we performed 100\,000 realizations. 

To quantify the uncertainties of the histograms of the synthetic sample, we perform a Monte-Carlo approach. Thus, we repeat the sampling procedure described above 700 times, and take, for each bin, the standard deviation of the resulting values as the uncertainty.

\subsection{Comparing the surface properties of the observed data and the simulated sample}
\begin{figure}
    \centering
    \includegraphics[width=\linewidth]{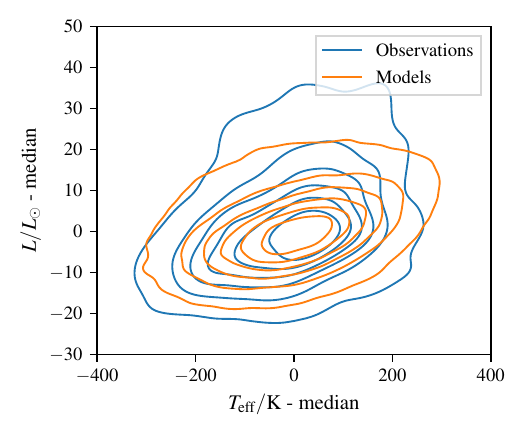}
    \caption{Kernel density estimations of the distributions of luminosities and effective temperature of the observed (blue) and synthetic (orange) samples, substracted by the median value of the sample.}
    \label{kde_hr}
\end{figure}
To verify that our sampling is compatible with the observed sample, we performed a simulation of the Hertzsprung-Russell diagram of our sample. To do so, we follow a similar methodology as the one described in Sect.~\ref{grid_sampling}, and perturb the obtained values of effective temperatures and luminosities by adding a gaussian uncertainty of $50\,\mathrm{K}$ and $5\,L_{\odot}$, respectively. Observational values of effective temperatures are taken from the APOGEE DR17 sample, and the luminosities from \cite{Berger2018}. The latter encompasses 2063 stars out of the 2110 stars of the observational sample used elsewhere in this work.

We present in Fig.~\ref{kde_hr} the Kernel Density Estimation (KDE) of the effective temperature and luminosities of the synthetic and observational samples. In order to correct for the systematic shift between the two, we subtracted the median of each sample. The differences between the median value of the synthetic and observed samples are $-164\,\mathrm{K}$ for the effective temperature and $-2.23\,L_\odot$ for the luminosity. The first may be attributed to the fact that we computed models with a fixed value of the mixing-length parameter, while the second is smaller than the typical observational uncertainties.

We find that the range of effective temperature and luminosities covered by our simulations is compatible with the one of the observations, hence showing the consistency between the synthetic and observed samples. We also find that the observed sample has a higher number of stars with high luminosities, but we note that most of these stars have a large ($> 10\,L\odot$) luminosity uncertainty.

\subsection{Computing period spacings}
\label{subsub_period_spacings}

The period spacings $\Delta \Pi$ of dipole modes are computed in this work using the asymptotic formulation of \cite{shibahashi79}, namely:

\begin{equation}
    \label{eq_period_spacing}
    \Delta \Pi = \sqrt{2} \pi^2 \left( \int_{r_0}^{r_1} \frac{N}{r}\,\dd r \right)^{-1},
\end{equation}
with $r_{0,1}$ being the boundaries of the g-mode resonating cavity, $N$ the
\brunt frequency, and $r$ the radial coordinate. 

The boundaries of the g-mode cavity are determined differently depending on whether mode trapping is taken into account or not. In the first case, $r_0$ is taken at the outermost radius of the maximal overshoot, overmixing or semi-convective region, depending on the CBM scheme. Indeed, there is at this position a steep helium discontinuity, which can reflect waves such that most of the oscillating energy of the observed modes is situated beyond $r_0$. Using such lower boundary allows us to compute a period spacing that is consistent with the frequencies computed using a stellar oscillation code (see Appendix~\ref{Appendix_frequencies}). Moreover, such approach is similar to the one used in \cite{Constantino2017}. We note that, in this ``mode-trapping'' scenario, the value of the period spacing is independent of the chemical and thermal stratification within the CBM region, but is still sensitive to the radial extent of the CBM region.

In the second scenario, we do not take into account mode trapping.
Therefore, we define $r_0$ as the inner boundary of the G-mode cavity, i.e. 
where, formally, $N^2 > \omega_{n,l}$, with $\omega_{n,l}$ being the angular 
frequency of the mode \citep{shibahashi79}. In our case, as the angular 
frequencies of the observed modes are situated around the angular frequency of 
maximum oscillation power $\omega_{\max}$, we define $r_0$ as the radius where 
$N^2 > \omega_{\max}$,. We compute $\omega_{\max}$ using the scaling relation $
\omega_{\max} = \omega_{\max, \odot} (M/M_{\odot})(R/R_{\odot})^{-2} 
(T_{\mathrm{eff}}/T_{\mathrm{eff},\odot})^{-0.5}$ \citep{Kjeldsen1995}. In that
case, contrary to the ``mode-trapping'' case, it is directly sensitive to the
stratification in temperature and composition inside the CBM region.
Finally, in both cases, $r_1$ is defined as the outermost radius where $N^2 > 
\omega_{\max}^2$. 

Typical Brunt-Väisälä profiles for the four different CBM scenarios are presented in Fig.~\ref{structure_models}. We indicate the region over which $N/r$ is integrated in the trapped scenario in red, in the non-trapped scenario in purple.

\section{Results}
\label{section_results}

In this section, we present first the period spacing distributions that we obtained for different core boundary mixing scenarios, assuming either the presence or the absence of mode trapping (respectively in Sect.~\ref{result_mode_trapping} and \ref{subsection_mixing_no_trapping}). Then, in Sect.~\ref{subsection_results_rates}, we investigate the effect of varying the rate of the $\reac$ reaction on the distributions. 

\subsection{In the presence of mode trapping}
\label{result_mode_trapping}

\begin{figure*}
    \centering
    \includegraphics[width=17cm]{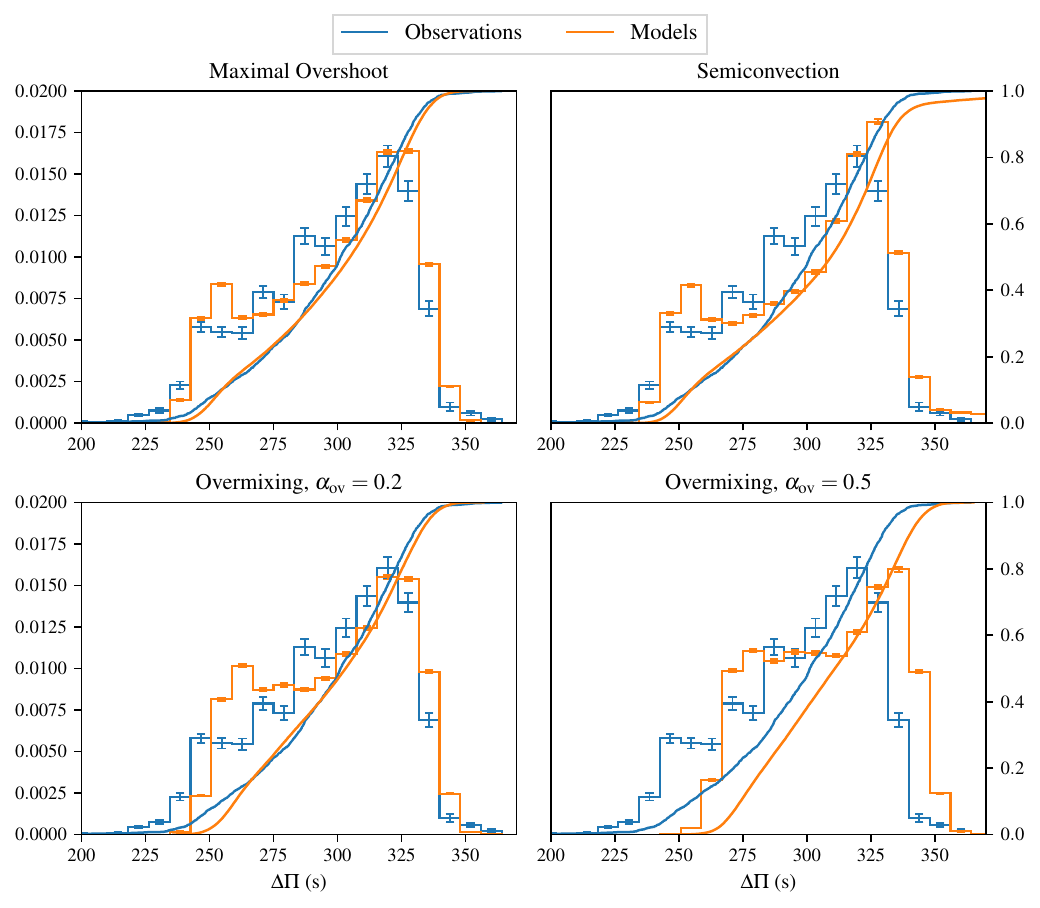}
    \caption{Distributions of the computed values of period spacing (orange) and of the observed values of period spacing (blue), for models that assume mode trapping. The lines represented the corresponding cumulative distributions. The simulated distributions are computed using a maximal overshoot scheme (upper left), a semiconvection scheme (upper right), an overmixing scheme with $\ov = 0.2$ (lower left) and $\ov = 0.5$ (lower right). The represented bin uncertainties for the observations are computed following the procedure explained in Appendix~\ref{uncertainty_bin}. Each bin value and associated uncertainties is normalized, i.e. divided by the total count and the bin width.}
    \label{results_mixing}
\end{figure*}

In this section, we compute the period spacing in the ``mode trapping'' scenario, as described in Section~\ref{subsub_period_spacings}. Fig.~\ref{results_mixing} shows the distributions of period spacing computed with 4 different CBM schemes.  The best fit models are the ones constructed by  using the maximal overshoot scheme, which is able to reproduce the main properties of the observed period spacing distributions. Yet, we can note that there is an overestimation of models with high values of period spacing ($> 325\,\mathrm{s}$), that can be resolved by decreasing the rate of the $\reac$ reaction (see Sect.~\ref{results_rates}). Also, we observe two other features that are not reproduced by the models:  a significant peak is present in the models around $250\,\mathrm{s}$  but not in the observations. The origin of this peak is discussed in Sect.~\ref{the_250s_peak}. Also, around $290\,\mathrm{s}$, an observational bin has a value significantly higher than the one predicted from the models. These two features are dominating the differences between the models and the observations. 

The semiconvection scheme yields a poorer fit, because of the CBPs that occur in these models. Indeed, CBPs lead to a significant number of realizations with very high values of period spacing ($400\,\mathrm{s}$, not visible on this plot) as well as a lower number of realizations around $300\,\mathrm{s}$, compared to the distribution resulting from maximal overshoot. Both of these features are not in line with the observed distributions.

Regarding the core extent, we only represent in Fig.~\ref{results_mixing} period
spacings computed using an overmixing scheme, for two values of $\ov$, 0.2 and
0.5. We can see that adding overshoot yield a poorer fit to the observations
compared to maximal overshoot. Notably, due to the extension of the core, the
predicted number of models with low values of period spacing (from 225 to
$250\,\mathrm{s}$) is significantly lower than the one observed in the
\emph{Kepler} data. Also, for models computed with a large value of $\ov$, there
is a strong overestimation of the number of stars with higher values of period
spacing. This is not the case for $\ov = 0.2$, where the predicted number of
stars at high values of period spacing is similar to the maximal overshoot case.
This is due to the fact that, during the late CHeB phase, the extent of the CBM
region (i.e., overshoot and semi-convection) of models with $\ov = 0.2$ is
similar to the extent of the fully mixed region of maximal overshoot models. 

\subsection{In the absence of mode trapping}
\label{subsection_mixing_no_trapping}

\begin{figure*}
    \centering
    \includegraphics[width=17cm]{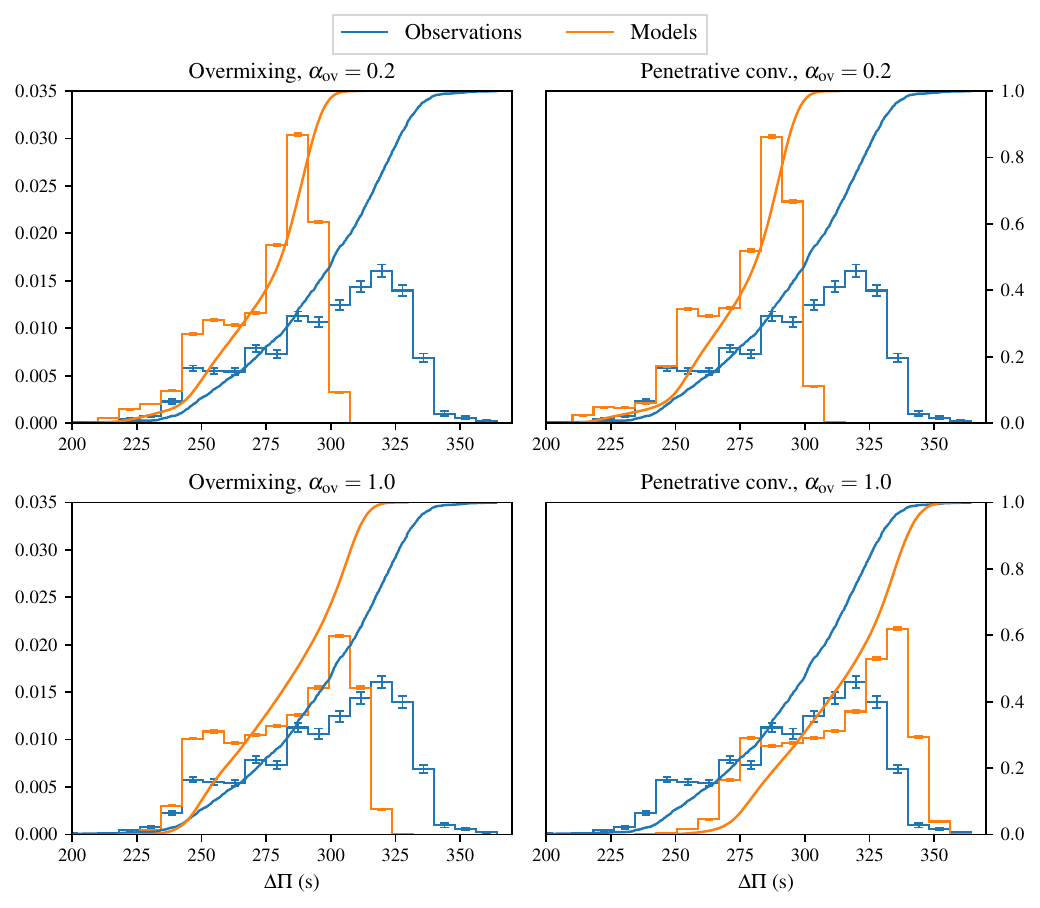}
    \caption{The same as in Fig.~\ref{results_mixing}, but without considering mode trapping, i.e. with the period spacings computed in the full g-mode cavity.}
    \label{results_mixing_no_trapping}
\end{figure*}

In this section, we compute the period spacing as in the ``non-trapped'' case, i.e. integrating over the full g-mode cavity region. The results of our computations are shown in Fig.~\ref{results_mixing_no_trapping}. We focus on the cases of overmixing and penetrative convection, as the results are similar for maximal overshoot models and deviates significantly from the observations for semiconvective models (see Fig.~9 of \citetalias{Noll2024}). A striking result is that, whatever the value of $\ov$ or the kind of temperature stratification in the overshoot region, none of these schemes can reproduce the observations if no mode trapping is assumed. In the case of overmixing, even with a high value of $\ov$, the simulated distributions underestimate the number of high period spacing models. Regarding penetrative convection, the distributions computed with a low or high value of $\ov$ are both incompatible with the highest and lowest values of observed period spacings. No ``sweet-spot'' value of $\ov$ allows us to produce a distribution that is compatible with the observations. 

A potential way to solve this discrepancy would be a temperature stratification that evolves during the CHeB phase, such that it is at first radiative and then adiabatic. Also, an evolving value of $\ov$ could improve the fit, if using penetrative convection. However, the study of such parametrization, which would lead to a much extended parameter space, is beyond the scope of this paper.

\subsection{$\reac$ Nuclear reaction rates}
\label{subsection_results_rates}
\begin{figure*}
    \centering
    \includegraphics[width=17cm]{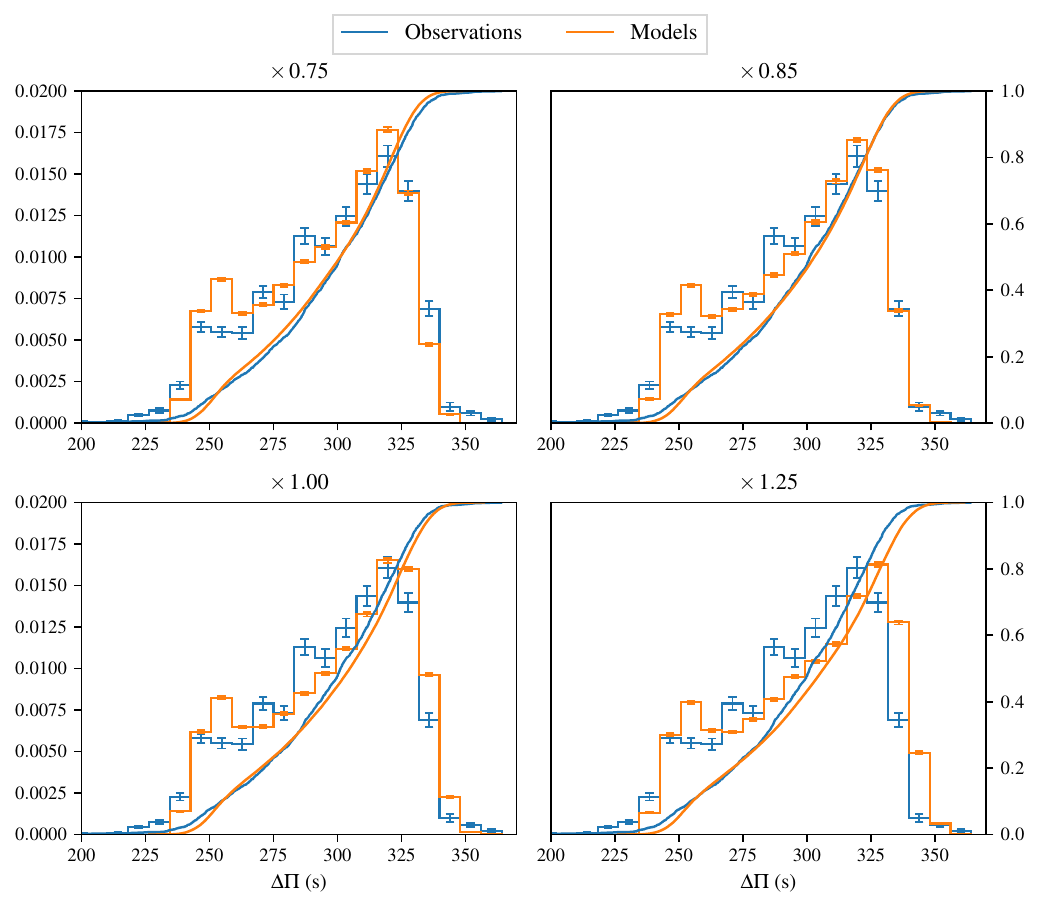}
    \caption{The same as in Fig.~\ref{results_mixing}, now for maximal overshoot and
    with varying $\reac$ nuclear reaction rates.}
    \label{results_rates}
\end{figure*}

For testing nuclear reaction rates,  we computed all our models using the maximal overshoot scheme as it is the one that is the most compatible with the observations. The rates are varied by multiplying the nominal value from \cite{Xu2013} by a given value. A comparison between the values of the different nuclear rates computed in this work and the recommendations from \cite{deBoer2017} are shown in Fig.~\ref{appendix_ratio_rates}. 

Fig.~\ref{results_rates} shows the computed distributions for different rates of the $\reac$ reaction. The effect is, as expected from \citetalias{Noll2024}, lower than the one resulting from varying the CBM prescription. We find that a higher rate leads to a larger number of predicted models with high values of period spacing. This is a consequence of the increased maximum value of period spacing reached by models with rate for $\reac$, as it lengthens the duration of the CHeB phase \citepalias[see][]{Noll2024}. Decreasing the rate has the opposite effect. We observe that it improves the quality of fit, especially for the higher values of period spacing. In particular, multiplying the rate of $\reac$ by $0.85$ seems to yield the closest distribution. As one can see in Fig.~\ref{appendix_ratio_rates}, this is approximately equivalent to the lower recommended rate of \cite{deBoer2017}. 

\section{The peak at 250\,s}
\label{the_250s_peak}

\begin{figure}
    \centering
    \includegraphics[width=9cm]{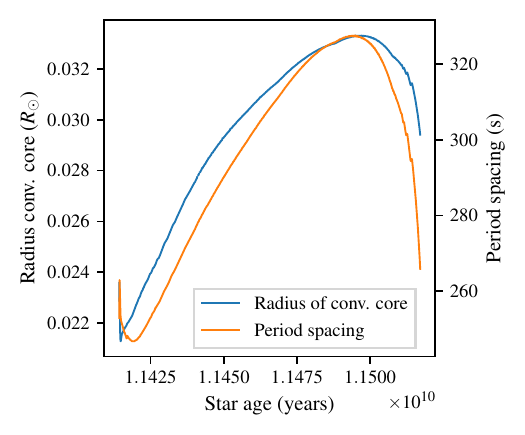}
    \caption{Evolution of the radius of the convective core (blue) and of the period spacing (orange) for a $1\,\smass$, maximal-overshoot model with solar metallicity, during the CHeB phase.} 
    \label{evol_core_period_spacing}
\end{figure}

\begin{figure}
    \centering
    \includegraphics[width=9cm]{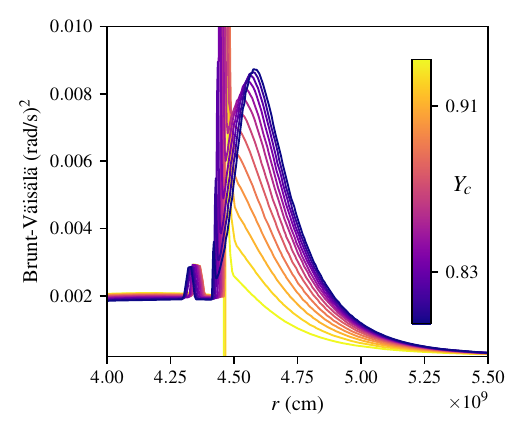}
    \caption{Evolution of the Brunt-Väisälä frequency profile at the location of the H-burning shell, at the very start of the CHeB phase. The lines are colored following the central helium composition: evolution goes from yellow to purple.}
    \label{evol_peak_N2}
\end{figure}

In the best-fit case, the feature that dominates the differences between modeled and observed distributions is a peak around $250\,\mathrm{s}$, visible for instance in the upper right panel of Fig.~\ref{results_rates}. As mentioned in \cite{Bossini2015}\footnote{Interestingly enough, the data distributions shown in Fig.~9 of \cite{Bossini2015}, which are taken from \cite{Mosser2014}, present a peak around 250 s. This peak is coincidental and caused by the smaller sample from \cite{Mosser2014}. Indeed, we find back this peak when plotting the period spacing of the \cite{Mosser2014} subsample using the results from \cite{Vrard2016}.}, this peak is due to the behavior of the period spacing at the very start of the CHeB phase. At the very beginning of the CHeB phase, $\Delta \Pi$ first decreases before continuously increasing until the final shrinking of the convective core, as illustrated in Fig.~\ref{evol_core_period_spacing}. Because of that initial decrease followed by an increase, the model spends more time with a period spacing around $250\,\mathrm{s}$, hence the overpopulation of models with such values of period spacing in the final simulated observations. 

We try next to understand the reason for such decrease at the start of the CHeB phase. It is not due to a variation in the mass of the core: as shown in Fig.~\ref{evol_core_period_spacing}, the decrease of the mass of the core at the beginning of the CHeB phase happens too quickly to be compatible with the period spacing variation. Rather, as mentioned in \cite{Constantino2015}, the decrease in period spacing is caused by the properties of the hydrogen-burning shell. To explore this, we show in Fig.~\ref{evol_peak_N2} the evolution of the profile of the Brunt-Väisälä frequency at the location of the H-burning shell, at the very beginning of the CHeB phase. The bump, situated from $4.5\times 10^9\,\mathrm{cm}$ outwards, is caused by the chemical gradient between the inner helium-rich region and the outer hydrogen-rich region. That bump is initially very narrow: this is a residual from the structure of the red giant branch, where the internal temperatures are high, and thus the energy production particularly localized. This leads to a narrow transition between the helium rich and hydrogen rich regions, and thus a strong chemical gradient.  However, during the CHeB phase, the temperatures are lower and therefore the H-burning occurs in a larger region. This leads to a widening and flattening of the chemical composition gradient and thus the Brunt-Väisälä frequency, at the start of the CHeB phase when the structure of the star adapts to the new burning phase. Consequently, the value of the integral in Eq.~\ref{eq_period_spacing} increases, leading to a decrease in period spacing.  Once the chemical gradient of the H-burning shell reaches its final shape, the variations of the period spacing are dominated by the evolution of the convective core: $\Delta \Pi$ increases. 

It is unclear why the $250\,\mathrm{s}$ peak does not appear in the observations. A selection effect due to peculiar global observables (such as luminosity, temperature or large separation) is unlikely, as the early RC stars do not have distinct values of these compared to more evolved RC stars. However, as mentioned in Sect.~3.3 of \cite{Constantino2015}, a potential bias could exist if the oscillation spectrum of these stars is messier than that of the other CHeB stars: it would be more difficult to measure the period spacing of a star with a messy power spectrum, which would automatically reduce their numbers in the sample of \cite{Vrard2016}. Also, one can note the work of \cite{Singh2021}, in which they measured the period spacing for super lithium-rich CHeB stars. They found several stars which are not part of the \cite{Vrard2016} sample and have a period spacing around 250\,s. 

Form a different perspective, the features of this peak can be influenced by the distribution in initial helium abundance, $Y_0$, of the sample. Indeed, as it can be seen in Fig.~8 of \citetalias{Noll2024}, modifying $Y_0$ has an impact on the value of the period spacing at the start of the CHeB phase, where the value of the period spacing is typically around $250\,\mathrm{s}$. In this work, we assumed a fixed enrichment law, with a slope of 2. A larger variety in the initial helium abundance may somewhat smooth out the $250\,\mathrm{s}$ peak. We propose exploring this question in a future work.

Finally, as the $250\,\mathrm{s}$ peak is caused by the sharp Brunt-Väisälä profile at the beginning of the CHeB phase, we checked if microscopic diffusion may help smooth it, but found that the effect is negligible and does not impact the computed distributions significantly. 

\section{Conclusions}
\label{section_conclusion}

In this work, we simulated the distribution of the period spacings of the RC stars of the \emph{Kepler} sample, in order to test the validity of the core boundary mixing and $\reac$ rates. To do so, for each core boundary mixing scheme or $\reac$ reaction rate, we computed a grid of CHeB tracks using the MESA stellar evolution code, with varying values of metallicities and masses. We then drew samples from these grids with priors that are as close as possible to the observed sample: for the metallicity and the masses, we used the spectroscopic and seismic values, respectively, from \cite{Pinsonneault2025}. For the age, we assumed that the stars are uniformly distributed along the CHeB phase. Finally, we perturbed the sampled period spacing values with uncertainties that are similar to the observational ones. This process allowed us to retrieve a period spacing distribution, that we compared to the observational one.

We found that we obtain the best fits when assuming a maximal overshoot mixing scheme. This scheme is, by definition, non-physical, yet, it gives period spacing values that are equal to the ones of a model with an induced semi-convective region, that does not exhibit any core breathing pulse and in which the observed modes are trapped outside the semi-convective region. The latter scenario, that is more physically justified, is our preferred interpretation of our results. We also found that adding overmixing, or penetrative convection, worsens the fit. Finally, we found that we cannot reproduce the observations if we compute the period spacings without taking into account the mode trapping, whatever mixing scheme is used. 

Regarding the rate of the $\reac$ reaction, decreasing the nominal \cite{Xu2013} by $15\%$ slightly improves the fit to the observed distribution, by decreasing the number of stars with high values of period spacing. Such decreased rate approximately corresponds to the lower recommended rate from \cite{deBoer2017}.

Finally, we stress that the model distributions cannot reproduce all the features of the observations. The main difference is the so-called 250\,s peak, that is an overpopulation of models with a period spacing around 250\,s that is not found in the observations. This overpopulation is the result of the decrease of the period spacing at the very start of the CHeB phase, due to the adaptation of the structure of the H-burning shell to the CHeB phase: thus, its presence is well understood and not due to numerical artifacts. It is therefore unclear why such overpopulation cannot be found in the observed sample: a possible explanation could be that the oscillation spectrum of the young RC stars that populate this 250\,s population, which just passed through the helium flash, is too messy to allow any clear measurement of the period spacing. Also, a very wide variety in initial helium abundance within the \emph{Kepler} sample could help ``smoothen'' the peak. Additional data, from e.g. the PLATO mission \citep{plato}, could shed new light on this issue. 

\begin{acknowledgements}
The authors thank the anonymous referee for their valuable comments which helped improve the discussion and the clarity of the paper. AN thanks Simon Campbell for interesting discussions about the 250\,s peak. AN and SH acknowledge funding from the ERC Consolidator Grant DipolarSound (grant agreement \#10s1000296). SB acknowledges NSF grant AST-2205026. AN also acknowledges funding from the program Unidad de Excelencia Maria de Maeztu (CEX2020-001058-M), the Generalitat de Catalunya (2021-SGR-1526) and the Tecnologías avanzadas para la exploración del universo project, within the framework of NextGenerationEU PRTR. 
\end{acknowledgements}

\bibliographystyle{aa}
\bibliography{biblio}

\begin{thebibliography}{51}
\expandafter\ifx\csname natexlab\endcsname\relax\def\natexlab#1{#1}\fi

\bibitem[{{Abdurro'uf} {et~al.}(2022){Abdurro'uf}, {Accetta}, {Aerts}, {Silva
  Aguirre}, {Ahumada}, {Ajgaonkar}, {Filiz Ak}, {Alam}, {Allende Prieto},
  {Almeida}, \& et~al.}]{Abdurro'uf2022}
{Abdurro'uf}, {Accetta}, K., {Aerts}, C., {et~al.} 2022, \apjs, 259, 35

\bibitem[{{Ahumada} {et~al.}(2020){Ahumada}, {Allende Prieto}, {Almeida},
  {Anders}, {Anderson}, {Andrews}, {Anguiano}, {Arcodia}, {Armengaud},
  {Aubert}, \& et~al.}]{Ahumada2020}
{Ahumada}, R., {Allende Prieto}, C., {Almeida}, A., {et~al.} 2020, \apjs, 249,
  3

\bibitem[{{Berger} {et~al.}(2018){Berger}, {Huber}, {Gaidos}, \& {van
  Saders}}]{Berger2018}
{Berger}, T.~A., {Huber}, D., {Gaidos}, E., \& {van Saders}, J.~L. 2018, \apj,
  866, 99

\bibitem[{{Borucki} {et~al.}(2010){Borucki}, {Koch}, {Basri}, {Batalha},
  {Brown}, {Caldwell}, {Caldwell}, {Christensen-Dalsgaard}, {Cochran},
  {DeVore}, {Dunham}, {Dupree}, {Gautier}, {Geary}, {Gilliland}, {Gould},
  {Howell}, {Jenkins}, {Kondo}, {Latham}, {Marcy}, {Meibom}, {Kjeldsen},
  {Lissauer}, {Monet}, {Morrison}, {Sasselov}, {Tarter}, {Boss}, {Brownlee},
  {Owen}, {Buzasi}, {Charbonneau}, {Doyle}, {Fortney}, {Ford}, {Holman},
  {Seager}, {Steffen}, {Welsh}, {Rowe}, {Anderson}, {Buchhave}, {Ciardi},
  {Walkowicz}, {Sherry}, {Horch}, {Isaacson}, {Everett}, {Fischer}, {Torres},
  {Johnson}, {Endl}, {MacQueen}, {Bryson}, {Dotson}, {Haas}, {Kolodziejczak},
  {Van Cleve}, {Chandrasekaran}, {Twicken}, {Quintana}, {Clarke}, {Allen},
  {Li}, {Wu}, {Tenenbaum}, {Verner}, {Bruhweiler}, {Barnes}, \&
  {Prsa}}]{borucki10}
{Borucki}, W.~J., {Koch}, D., {Basri}, G., {et~al.} 2010, Science, 327, 977

\bibitem[{{Bossini} {et~al.}(2015){Bossini}, {Miglio}, {Salaris},
  {Pietrinferni}, {Montalb{\'a}n}, {Bressan}, {Noels}, {Cassisi}, {Girardi}, \&
  {Marigo}}]{Bossini2015}
{Bossini}, D., {Miglio}, A., {Salaris}, M., {et~al.} 2015, \mnras, 453, 2290

\bibitem[{{Bossini} {et~al.}(2017){Bossini}, {Miglio}, {Salaris}, {Vrard},
  {Cassisi}, {Mosser}, {Montalb{\'a}n}, {Girardi}, {Noels}, {Bressan},
  {Pietrinferni}, \& {Tayar}}]{Bossini2017}
{Bossini}, D., {Miglio}, A., {Salaris}, M., {et~al.} 2017, \mnras, 469, 4718

\bibitem[{{Bressan} {et~al.}(1986){Bressan}, {Bertelli}, \&
  {Chiosi}}]{Bressan1986}
{Bressan}, A., {Bertelli}, G., \& {Chiosi}, C. 1986, \memsai, 57, 411

\bibitem[{{Caputo} {et~al.}(1989){Caputo}, {Castellani}, {Chieffi}, {Pulone},
  \& {Tornambe}}]{Caputo1989}
{Caputo}, F., {Castellani}, V., {Chieffi}, A., {Pulone}, L., \& {Tornambe}, A.,
  J. 1989, \apj, 340, 241

\bibitem[{{Castellani} {et~al.}(1971{\natexlab{a}}){Castellani}, {Giannone}, \&
  {Renzini}}]{Castellani1971b}
{Castellani}, V., {Giannone}, P., \& {Renzini}, A. 1971{\natexlab{a}}, \apss,
  10, 355

\bibitem[{{Castellani} {et~al.}(1971{\natexlab{b}}){Castellani}, {Giannone}, \&
  {Renzini}}]{Castellani1971}
{Castellani}, V., {Giannone}, P., \& {Renzini}, A. 1971{\natexlab{b}}, \apss,
  10, 340

\bibitem[{{Chidester} {et~al.}(2022){Chidester}, {Farag}, \&
  {Timmes}}]{Chidester2022}
{Chidester}, M.~T., {Farag}, E., \& {Timmes}, F.~X. 2022, \apj, 935, 21

\bibitem[{{Constantino} {et~al.}(2015){Constantino}, {Campbell},
  {Christensen-Dalsgaard}, {Lattanzio}, \& {Stello}}]{Constantino2015}
{Constantino}, T., {Campbell}, S.~W., {Christensen-Dalsgaard}, J., {Lattanzio},
  J.~C., \& {Stello}, D. 2015, \mnras, 452, 123

\bibitem[{{Constantino} {et~al.}(2017){Constantino}, {Campbell}, \&
  {Lattanzio}}]{Constantino2017}
{Constantino}, T., {Campbell}, S.~W., \& {Lattanzio}, J.~C. 2017, \mnras, 472,
  4900

\bibitem[{{Constantino} {et~al.}(2016){Constantino}, {Campbell}, {Lattanzio},
  \& {van Duijneveldt}}]{Constantino2016}
{Constantino}, T., {Campbell}, S.~W., {Lattanzio}, J.~C., \& {van Duijneveldt},
  A. 2016, \mnras, 456, 3866

\bibitem[{{Cyburt} {et~al.}(2010){Cyburt}, {Amthor}, {Ferguson}, {Meisel},
  {Smith}, {Warren}, {Heger}, {Hoffman}, {Rauscher}, {Sakharuk}, {Schatz},
  {Thielemann}, \& {Wiescher}}]{Cyburt2010}
{Cyburt}, R.~H., {Amthor}, A.~M., {Ferguson}, R., {et~al.} 2010, \apjs, 189,
  240

\bibitem[{{deBoer} {et~al.}(2017){deBoer}, {G{\"o}rres}, {Wiescher}, {Azuma},
  {Best}, {Brune}, {Fields}, {Jones}, {Pignatari}, {Sayre}, {Smith}, {Timmes},
  \& {Uberseder}}]{deBoer2017}
{deBoer}, R.~J., {G{\"o}rres}, J., {Wiescher}, M., {et~al.} 2017, Reviews of
  Modern Physics, 89, 035007

\bibitem[{{Dorman} \& {Rood}(1993)}]{Dorman1993}
{Dorman}, B. \& {Rood}, R.~T. 1993, \apj, 409, 387

\bibitem[{{Gabriel} {et~al.}(2014){Gabriel}, {Noels}, {Montalb{\'a}n}, \&
  {Miglio}}]{gabriel14}
{Gabriel}, M., {Noels}, A., {Montalb{\'a}n}, J., \& {Miglio}, A. 2014, \aap,
  569, A63

\bibitem[{{Girardi}(2016)}]{Girardi2016}
{Girardi}, L. 2016, \araa, 54, 95

\bibitem[{{Grevesse} \& {Sauval}(1998)}]{gs98}
{Grevesse}, N. \& {Sauval}, A.~J. 1998, \ssr, 85, 161

\bibitem[{{Iglesias} \& {Rogers}(1996)}]{opal_opacities}
{Iglesias}, C.~A. \& {Rogers}, F.~J. 1996, \apj, 464, 943

\bibitem[{{Irwin}(2012)}]{Irwin2012}
{Irwin}, A.~W. 2012, {FreeEOS: Equation of State for stellar interiors
  calculations}, Astrophysics Source Code Library, record ascl:1211.002

\bibitem[{{Jermyn} {et~al.}(2023){Jermyn}, {Bauer}, {Schwab}, {Farmer}, {Ball},
  {Bellinger}, {Dotter}, {Joyce}, {Marchant}, {Mombarg}, {Wolf}, {Sunny Wong},
  {Cinquegrana}, {Farrell}, {Smolec}, {Thoul}, {Cantiello}, {Herwig}, {Toloza},
  {Bildsten}, {Townsend}, \& {Timmes}}]{Jermyn2023}
{Jermyn}, A.~S., {Bauer}, E.~B., {Schwab}, J., {et~al.} 2023, \apjs, 265, 15

\bibitem[{{Jermyn} {et~al.}(2021){Jermyn}, {Schwab}, {Bauer}, {Timmes}, \&
  {Potekhin}}]{Jermyn2021}
{Jermyn}, A.~S., {Schwab}, J., {Bauer}, E., {Timmes}, F.~X., \& {Potekhin},
  A.~Y. 2021, \apj, 913, 72

\bibitem[{{Kjeldsen} \& {Bedding}(1995)}]{Kjeldsen1995}
{Kjeldsen}, H. \& {Bedding}, T.~R. 1995, \aap, 293, 87

\bibitem[{{Kuhfuss}(1986)}]{Kuhfuss1986}
{Kuhfuss}, R. 1986, \aap, 160, 116

\bibitem[{{Kunz} {et~al.}(2002){Kunz}, {Fey}, {Jaeger}, {Mayer}, {Hammer},
  {Staudt}, {Harissopulos}, \& {Paradellis}}]{Kunz2002}
{Kunz}, R., {Fey}, M., {Jaeger}, M., {et~al.} 2002, \apj, 567, 643

\bibitem[{{Langer} {et~al.}(1983){Langer}, {Fricke}, \&
  {Sugimoto}}]{Langer1983}
{Langer}, N., {Fricke}, K.~J., \& {Sugimoto}, D. 1983, \aap, 126, 207

\bibitem[{{Mehta} {et~al.}(2022){Mehta}, {Buonanno}, {Gair}, {Miller}, {Farag},
  {deBoer}, {Wiescher}, \& {Timmes}}]{Mehta2022}
{Mehta}, A.~K., {Buonanno}, A., {Gair}, J., {et~al.} 2022, \apj, 924, 39

\bibitem[{{Montalb{\'a}n} {et~al.}(2013){Montalb{\'a}n}, {Miglio}, {Noels},
  {Dupret}, {Scuflaire}, \& {Ventura}}]{Montalban2013}
{Montalb{\'a}n}, J., {Miglio}, A., {Noels}, A., {et~al.} 2013, \apj, 766, 118

\bibitem[{{Mosser} {et~al.}(2014){Mosser}, {Benomar}, {Belkacem}, {Goupil},
  {Lagarde}, {Michel}, {Lebreton}, {Stello}, {Vrard}, {Barban}, {Bedding},
  {Deheuvels}, {Chaplin}, {De Ridder}, {Elsworth}, {Montalban}, {Noels},
  {Ouazzani}, {Samadi}, {White}, \& {Kjeldsen}}]{Mosser2014}
{Mosser}, B., {Benomar}, O., {Belkacem}, K., {et~al.} 2014, \aap, 572, L5

\bibitem[{{Mosser} {et~al.}(2012){Mosser}, {Goupil}, {Belkacem}, {Michel},
  {Stello}, {Marques}, {Elsworth}, {Barban}, {Beck}, {Bedding}, {De Ridder},
  {Garc{\'\i}a}, {Hekker}, {Kallinger}, {Samadi}, {Stumpe}, {Barclay}, \&
  {Burke}}]{Mosser2012}
{Mosser}, B., {Goupil}, M.~J., {Belkacem}, K., {et~al.} 2012, \aap, 540, A143

\bibitem[{{Noll} {et~al.}(2024){Noll}, {Basu}, \& {Hekker}}]{Noll2024}
{Noll}, A., {Basu}, S., \& {Hekker}, S. 2024, \aap, 683, A189

\bibitem[{{Noll} \& {Deheuvels}(2023)}]{Noll2023}
{Noll}, A. \& {Deheuvels}, S. 2023, \aap, 676, A70

\bibitem[{{Paxton} {et~al.}(2011){Paxton}, {Bildsten}, {Dotter}, {Herwig},
  {Lesaffre}, \& {Timmes}}]{Paxton2011}
{Paxton}, B., {Bildsten}, L., {Dotter}, A., {et~al.} 2011, \apjs, 192, 3

\bibitem[{{Paxton} {et~al.}(2013){Paxton}, {Cantiello}, {Arras}, {Bildsten},
  {Brown}, {Dotter}, {Mankovich}, {Montgomery}, {Stello}, {Timmes}, \&
  {Townsend}}]{Paxton2013}
{Paxton}, B., {Cantiello}, M., {Arras}, P., {et~al.} 2013, \apjs, 208, 4

\bibitem[{{Paxton} {et~al.}(2015){Paxton}, {Marchant}, {Schwab}, {Bauer},
  {Bildsten}, {Cantiello}, {Dessart}, {Farmer}, {Hu}, {Langer}, {Townsend},
  {Townsley}, \& {Timmes}}]{paxton15}
{Paxton}, B., {Marchant}, P., {Schwab}, J., {et~al.} 2015, \apjs, 220, 15

\bibitem[{{Paxton} {et~al.}(2018){Paxton}, {Schwab}, {Bauer}, {Bildsten},
  {Blinnikov}, {Duffell}, {Farmer}, {Goldberg}, {Marchant}, {Sorokina},
  {Thoul}, {Townsend}, \& {Timmes}}]{paxton18}
{Paxton}, B., {Schwab}, J., {Bauer}, E.~B., {et~al.} 2018, \apjs, 234, 34

\bibitem[{{Paxton} {et~al.}(2019){Paxton}, {Smolec}, {Schwab}, {Gautschy},
  {Bildsten}, {Cantiello}, {Dotter}, {Farmer}, {Goldberg}, {Jermyn}, {Kanbur},
  {Marchant}, {Thoul}, {Townsend}, {Wolf}, {Zhang}, \& {Timmes}}]{Paxton2019}
{Paxton}, B., {Smolec}, R., {Schwab}, J., {et~al.} 2019, \apjs, 243, 10

\bibitem[{{Pinsonneault} {et~al.}(2025){Pinsonneault}, {Zinn}, {Tayar},
  {Serenelli}, {Garc{\'\i}a}, {Mathur}, {Vrard}, {Elsworth}, {Mosser},
  {Stello}, {Bell}, {Bugnet}, {Corsaro}, {Gaulme}, {Hekker}, {Hon}, {Huber},
  {Kallinger}, {Cao}, {Johnson}, {Liagre}, {Patton}, {Santos}, {Basu}, {Beck},
  {Beers}, {Chaplin}, {Cunha}, {Frinchaboy}, {Girardi}, {Godoy-Rivera},
  {Holtzman}, {J{\"o}nsson}, {M{\'e}sz{\'a}ros}, {Reyes}, {Rix}, {Shetrone},
  {Smith}, {Spoo}, {Stassun}, \& {Wang}}]{Pinsonneault2025}
{Pinsonneault}, M.~H., {Zinn}, J.~C., {Tayar}, J., {et~al.} 2025, \apjs, 276,
  69

\bibitem[{{Rauer} {et~al.}(2014){Rauer}, {Catala}, {Aerts}, {Appourchaux},
  {Benz}, {Brandeker}, {Christensen-Dalsgaard}, {Deleuil}, {Gizon}, {Goupil},
  {G{\"u}del}, {Janot-Pacheco}, {Mas-Hesse}, {Pagano}, {Piotto}, {Pollacco},
  {Santos}, {Smith}, {Su{\'a}rez}, {Szab{\'o}}, {Udry}, {Adibekyan}, {Alibert},
  {Almenara}, {Amaro-Seoane}, {Eiff}, {Asplund}, {Antonello}, {Barnes},
  {Baudin}, {Belkacem}, {Bergemann}, {Bihain}, {Birch}, {Bonfils}, {Boisse},
  {Bonomo}, {Borsa}, {Brand {\~a}o}, {Brocato}, {Brun}, {Burleigh}, {Burston},
  {Cabrera}, {Cassisi}, {Chaplin}, {Charpinet}, {Chiappini}, {Church},
  {Csizmadia}, {Cunha}, {Damasso}, {Davies}, {Deeg}, {D{\'\i}az}, {Dreizler},
  {Dreyer}, {Eggenberger}, {Ehrenreich}, {Eigm{\"u}ller}, {Erikson}, {Farmer},
  {Feltzing}, {de Oliveira Fialho}, {Figueira}, {Forveille}, {Fridlund},
  {Garc{\'\i}a}, {Giommi}, {Giuffrida}, {Godolt}, {Gomes da Silva}, {Granzer},
  {Grenfell}, {Grotsch-Noels}, {G{\"u}nther}, {Haswell}, {Hatzes},
  {H{\'e}brard}, {Hekker}, {Helled}, {Heng}, {Jenkins}, {Johansen},
  {Khodachenko}, {Kislyakova}, {Kley}, {Kolb}, {Krivova}, {Kupka}, {Lammer},
  {Lanza}, {Lebreton}, {Magrin}, {Marcos-Arenal}, {Marrese}, {Marques},
  {Martins}, {Mathis}, {Mathur}, {Messina}, {Miglio}, {Montalban}, {Montalto},
  {Monteiro}, {Moradi}, {Moravveji}, {Mordasini}, {Morel}, {Mortier},
  {Nascimbeni}, {Nelson}, {Nielsen}, {Noack}, {Norton}, {Ofir}, {Oshagh},
  {Ouazzani}, {P{\'a}pics}, {Parro}, {Petit}, {Plez}, {Poretti}, {Quirrenbach},
  {Ragazzoni}, {Raimondo}, {Rainer}, {Reese}, {Redmer}, {Reffert},
  {Rojas-Ayala}, {Roxburgh}, {Salmon}, {Santerne}, {Schneider}, {Schou},
  {Schuh}, {Schunker}, {Silva-Valio}, {Silvotti}, {Skillen}, {Snellen}, {Sohl},
  {Sousa}, {Sozzetti}, {Stello}, {Strassmeier}, {{\v{S}}vanda}, {Szab{\'o}},
  {Tkachenko}, {Valencia}, {Van Grootel}, {Vauclair}, {Ventura}, {Wagner},
  {Walton}, {Weingrill}, {Werner}, {Wheatley}, \& {Zwintz}}]{plato}
{Rauer}, H., {Catala}, C., {Aerts}, C., {et~al.} 2014, Experimental Astronomy,
  38, 249

\bibitem[{{Salaris} {et~al.}(1993){Salaris}, {Chieffi}, \&
  {Straniero}}]{Salaris1993}
{Salaris}, M., {Chieffi}, A., \& {Straniero}, O. 1993, \apj, 414, 580

\bibitem[{{Schwarzschild} \& {Härm}(1969)}]{Schwarzschild1969}
{Schwarzschild}, M. \& {Härm}, R. 1969, in \baas, Vol.~1, 99

\bibitem[{{Shen} {et~al.}(2023){Shen}, {Guo}, {deBoer}, {Li}, {Li}, {Li},
  {Tang}, {Pang}, {Adhikari}, {Basu}, {Su}, {Yan}, {Fan}, {Liu}, {Chen}, {Han},
  {Li}, {Lian}, {Ma}, {Nan}, {Nan}, {Wang}, {Zeng}, {Zhang}, \&
  {Liu}}]{Shen2023}
{Shen}, Y., {Guo}, B., {deBoer}, R.~J., {et~al.} 2023, \apj, 945, 41

\bibitem[{{Shibahashi}(1979)}]{shibahashi79}
{Shibahashi}, H. 1979, \pasj, 31, 87

\bibitem[{{Singh} {et~al.}(2021){Singh}, {Reddy}, {Campbell}, {Kumar}, \&
  {Vrard}}]{Singh2021}
{Singh}, R., {Reddy}, B.~E., {Campbell}, S.~W., {Kumar}, Y.~B., \& {Vrard}, M.
  2021, \apjl, 913, L4

\bibitem[{{Spruit}(2015)}]{Spruit2015}
{Spruit}, H.~C. 2015, \aap, 582, L2

\bibitem[{{Sweigart} \& {Demarque}(1972)}]{Sweigart1972}
{Sweigart}, A.~V. \& {Demarque}, P. 1972, \aap, 20, 445

\bibitem[{{Townsend} \& {Teitler}(2013)}]{Townsend2013}
{Townsend}, R.~H.~D. \& {Teitler}, S.~A. 2013, \mnras, 435, 3406

\bibitem[{{Vrard} {et~al.}(2016){Vrard}, {Mosser}, \& {Samadi}}]{Vrard2016}
{Vrard}, M., {Mosser}, B., \& {Samadi}, R. 2016, \aap, 588, A87

\bibitem[{{Xu} {et~al.}(2013){Xu}, {Takahashi}, {Goriely}, {Arnould}, {Ohta},
  \& {Utsunomiya}}]{Xu2013}
{Xu}, Y., {Takahashi}, K., {Goriely}, S., {et~al.} 2013, \nphysa, 918, 61

\end{thebibliography}

\begin{appendix}
\section{Computation of the uncertainties of bins for observational data}
\label{uncertainty_bin}

In the histograms representing observational data, we took into account the observational uncertainties for the computation of the bin uncertainties, by computing them as follows. 

For each bin $j$, each observation $i$ can be considered as an independant Bernouilli trial, as the observation can be inside (``success'') or outside (``failure'') the bin bounds. We model the observed uncertainties as a normal distribution, $\mathcal{N}(x_i, \sigma_i)$ with $x_i$ the observed value and $\sigma_i$ the associated uncertainty. Then, the probability that the observation $i$ is inside the bin $j$ is:
\begin{equation}
    p_i(j) = \int_{l_j}^{u_j} \frac{1}{\sigma_i \sqrt{2\pi}}  \exp \left[ -\frac{1}{2} \left(\frac{x - x_i}{\sigma_i} \right)^2 \right] \, \dd x,
\end{equation}
with $l_j$ and $u_j$ the lower and upper bounds of the bin $j$, respectively.  Therefore, the distribution of the bin value follows a Poisson binomial distribution, whose mean and variances are:
\begin{align}
    \mathrm{E}(X_j) &= \sum_i^N p_i(j), \\
    \sigma_j^2 & =  \sum_i^N p_i(j) (1-p_i(j)).
\end{align}

The latter is used as the uncertainty of the bin value. 

\section{Equivalence between the seismic properties of maximal overshoot and semiconvection models}
\label{appendix_eq_mo_sc}

\begin{figure}[h]
    \centering
    \includegraphics[width=9cm]{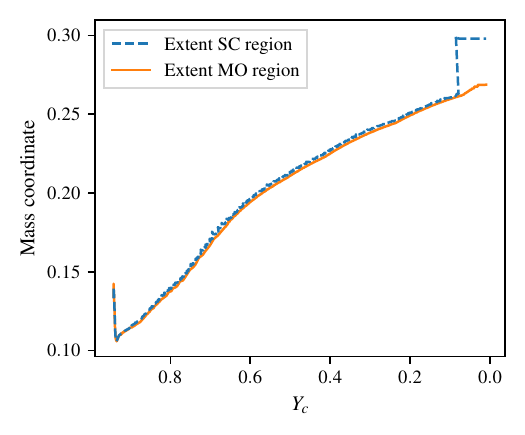}
    \caption{Evolution of the mass of the fully mixed core and semi convective region (blue) and of the maximal overshoot region (orange).}
    \label{fig_extent_sc}
\end{figure}

In this work, we use maximal overshoot (MO) models as seismic equivalent models
of a more physical scenario in which the modes are trapped outside an
semi-convective (SC) region. We can justify this by the fact that the extent of
the fully-mixed MO core is the same, during the CHeB phase, as the extent of the
CBM region for the SC models (i.e., the fully mixed core and the SC region). In
Fig.~\ref{fig_extent_sc}, we represent the extent of the fully-mixed core for MO
models, and of the CBM region for the SC models and indeed see that
they are similar\footnote{An equivalent plot can be found in \cite{Paxton2019},
Fig.~44.}. One key difference, however, is that the SC model exhibits core
breathing pulses (CBP), which are sudden increase of the core size at the end of
the CHeB phase. In Fig.~\ref{fig_extent_sc}, a core breathing pulse event can be
seen at $Y_{\mathrm{c}} = 0.1$. Such increase of the core mass lead to very high
values of period spacing, which are incompatible with the observations (see
Sect.~\ref{result_mode_trapping}). We can note that \cite{Spruit2015} raised an
argument, based on the higher buoyancy of helium compared to carbon and oxygen,
which limits the growth rate of the core and therefore inhibits the CBPs, which
has been confirmed by the models of \cite{Constantino2017}.

One could wonder why the SC region extent is very similar to the MO, despite the differences between the two schemes. The reason is that both require the local minimum of the radiative gradient in the core to be equal to the adiabatic gradient. This condition alone determines the evolution of the size of the mixed region: therefore, the extent of both SC and MO regions evolve similarly. Yet, we can note that two models with these schemes are not strictly equivalent: as the fully mixed region is larger in the MO case, the duration of the CHeB phase is extended for MO models. This, however, does not impact our work, as we sampled the parameter of space using the normalized $\tau$ variable rather than the absolute age. 

\section{Comparison between the \cite{Xu2013} and \cite{deBoer2017} rates for
the $\reac$ reaction}

\begin{figure}[h]
    \centering
    \includegraphics[width=9cm]{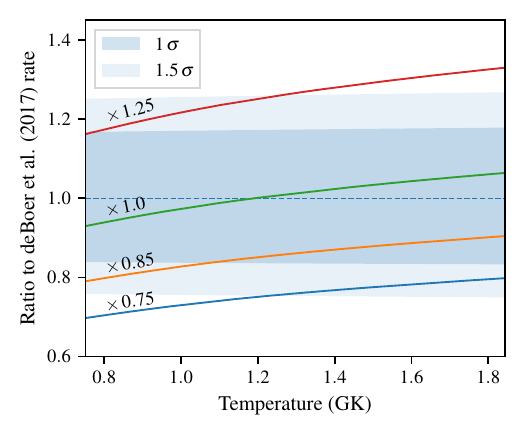}
    \caption{Ratio between the \cite{Xu2013} rates, multiplied by a given factor as done in Sect.~\ref{results_rates}, and the recommended rates from \cite{deBoer2017}, with their uncertainties indicated by the blue regions. Temperatures are typical of a core of a CHeB star. To represent the \cite{deBoer2017} rates, we used the tables with update temperature resolution from \cite{Mehta2022} and accessible through \cite{Chidester2022}. To compute the \cite{Xu2013} rates, we used the JINA Reaclib equation \citep{Cyburt2010}.}
    \label{appendix_ratio_rates}
\end{figure}

\section{Computing the frequencies and eigenfunctions}
\label{Appendix_frequencies}
\begin{figure}[h]
    \centering
    \includegraphics[width=9cm]{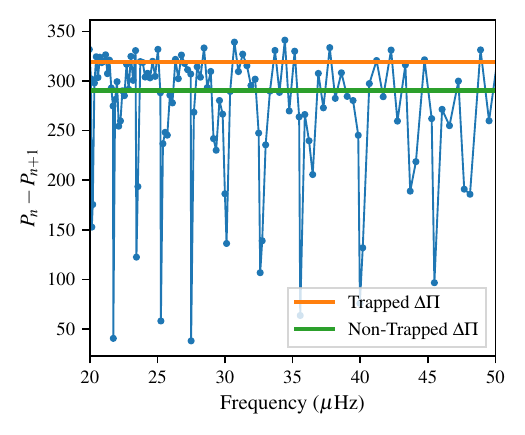}
    \caption{Consecutive period differences for a model computed with an overmixing scheme, terminated at $Y_{\mathrm{c}} = 0.35$. The asymptotic period spacing are represented by the horizontal lines, computed either in the non-trapping scenario (green) or in in the trapped scenario (orange). }
    \label{dpi_freq}
\end{figure}

In this work, we computed the period spacing using Eq.~\ref{eq_period_spacing}, assuming the inner boundary of the g-mode cavity either to be the outer boundary of the CBM region (``trapped'' scenario), or the outer boundary of the fully-mixed region (``non-trapped'' scenario) (see Fig.~\ref{structure_models} for an illustration of theses boundaries for different mixing scenarios). In the following, we compare the resulting, asymptotic period spacings to the frequencies computed with an oscillation code, GYRE \citep{Townsend2013}. We take the peculiar case of an overmixing model that is evolved enough to have a semi-convective region around the overmixing region, with a structure that is similar to the lower left panel of Fig.~\ref{structure_models}. We present the consecutive period spacing of this model in Fig.~\ref{dpi_freq}. As noted in \cite{Constantino2015}, the consecutive period spacing is quite chaotic due to the complex structure of the model, which has several discontinuities notably caused by the helium sub-flashes. Thus, it is difficult to clearly determine a period spacing out of it. Yet, the period spacing computed in the trapped case is in better agreement than the one computed in the non-trapped case. We note that the model used to computed these frequencies has been smoothed compared to the ones used in the rest of the work, in order to improve the regularity of the consecutive period spacing.

\begin{figure}[h]
    \centering
    \includegraphics[width=9cm]{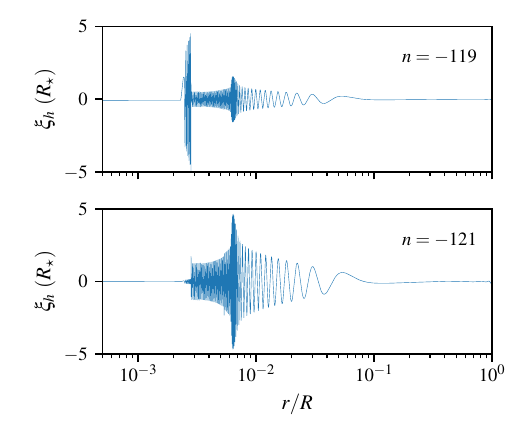}
    \caption{Horizontal displacement of dipolar modes with radial orders -119 and -121.}
    \label{fig_displacement}
\end{figure}

Moreover, we computed the eigenfunctions of two of the modes presented in Fig.~\ref{dpi_freq}, to investigate the properties of the trapped modes. We present in Fig.~\ref{fig_displacement} the horizontal displacement of the modes with radial orders $-119$ and $-121$. One can see that the mode $n=-119$ is mainly oscillating in the lower part of the cavity, i.e. the overshoot and the semi-convective region, and is therefore trapped in this region, while having a lower amplitude in the rest of the g-mode cavity. Oppositely, most of the modes (such as $n=-121$) have a small amplitude in the overshooting/semi-convection region and a large amplitude elsewhere in the g-mode cavity.

\end{appendix}

\end{document}